\documentclass{nature}
\usepackage[T1]{fontenc}
\usepackage[left]{lineno}
\usepackage{graphicx}
\usepackage{pdflscape}
\usepackage{hyperref}
\usepackage{threeparttable}
\usepackage{xcolor}
\usepackage{ulem}

\usepackage{caption}
\usepackage[caption = false]{subfig}
\usepackage{academicons}
\usepackage{amsmath, amssymb}

\newcommand{\apj}{Astrophys. J.}
\newcommand{\apjl}{Astrophys. J.}
\newcommand{\apjs}{Astrophys. J.}
\newcommand{\aap}{Astron. Astrophys.}
\newcommand{\mnras}{Mon. Not. R. Astron. Soc.}
\newcommand{\nat}{Nature}

\definecolor{orcidlogocol}{HTML}{A6CE39}

\def\be{\begin{eqnarray}}
\def\ee{\end{eqnarray}}

\makeatletter

\let\saved@includegraphics\includegraphics
\AtBeginDocument{\let\includegraphics\saved@includegraphics}
\renewenvironment*{figure}{\@float{figure}}{\end@float}
\makeatother

\makeatletter
\def\@fnsymbol#1{\ensuremath{\ifcase#1\or \dagger\or \ddagger\or
 \mathsection\or \mathparagraph\or \|\or **\or \dagger\dagger
 \or \ddagger\ddagger \else\@ctrerr\fi}}
\makeatother

\usepackage[utf8]{inputenc}
\usepackage{hyperref}
\usepackage{graphicx}
\usepackage{longtable}
\usepackage{supertabular}
\usepackage{float} 

\usepackage{aas_macros}

\newcommand{\target}{FRB~20240114A}
\newcommand{\arcdeg}{\mbox{$^{\circ}$}}

\newcommand{\fermi}{\textit{Fermi}}
\newcommand{\gr}{$\gamma$-ray}

\title{Flaring gamma-ray emission coincident with a hyperactive fast radio burst source}

\author{
Y. Xing$^{1}$,  
W. Yu$^{1}$\href{https://orcid.org/0000-0002-3844-9677}{\includegraphics[scale=0.08]{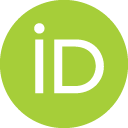}}\thanks{E-mail: wenfei@shao.ac.cn}, 
Z. Yan$^{1}$, X. Zhang$^{1}$, 
and B. Zhang$^{2,3}$\thanks{E-mail: zhang@physics.unlv.edu}}
\begin{document}
\maketitle
\begin{affiliations}

 \item Shanghai Astronomical Observatory, Chinese Academy of Sciences, Shanghai 200030, China
 \item The Nevada Center for Astrophysics, University of Nevada, Las Vegas, Las Vegas, NV 89154, USA
 \item Department of Physics and Astronomy, University of Nevada, Las Vegas, Las Vegas, NV 89154, USA 
 \end{affiliations}

\bigskip

\begin{abstract} 
Fast radio bursts (FRBs) are bright milliseconds-duration radio bursts from cosmological distances \cite{lorimer07,thornton13}. Despite intense observational and theoretical studies, their physical origin is still mysterious \cite{zhang23}. One major obstacle is the lack of identification of multi-wavelength counterparts for FRBs at cosmological distances. So far, all the searches other than in the radio wavelength \cite{2017ApJ...846...80S}, including those in the \gr\ energies \cite{ZhangARNPS24,2023A&A...675A..99P}, have only left upper limits. Here we report a gigaelectronvolt (GeV) \gr \ flare lasting 15.6 seconds as well as additional evidence of variable gamma-ray emission in temporal and spatial association with the hyper-active, newly discovered repeating \target, which has been localized to a dwarf galaxy at a redshift of 0.13 \cite{2024ATel16613....1B}. The energetic, short GeV \gr\ flare reached a prompt isotropic luminosity of the order of ${10}^{48}~{\rm ergs~{s}^{-1}}$. The additional less-significant \gr\ flares, if true, also have similar luminosities; such flares could contribute to a 5-day average luminosity of the order of ${10}^{45}~{\rm ergs~{s}^{-1}}$. These high-luminosity flares challenge the traditional FRB engine scenario involving a seconds-period magnetar. Rather, it suggests a powerful, long-lived but newborn energy source at the location of this active repeater, either directly powering the bursts or indirectly triggering bursts in the vicinity of the FRB engine. 



\end{abstract}

\smallskip

The repeating 
\target~was discovered by the {\it Canadian Hydrogen Intensity Mapping Experiment (CHIME)} on 2024 Jan 14. With only a 4-minute daily exposure window, the detection of multiple bursts by {\it CHIME}
in late January indicated that the FRB source had a high burst rate $\sim 15$ per hour for bursts brighter than 1 Jy ms in the 400--800 MHz band \cite{2024ATel16420....1S}. Since then, follow-up observations in the next few months with both large and small single dish radio telescopes and radio arrays have confirmed that \target~had been hyper-active, showing both extremely high burst rates up to the order of 500 bursts per hour with the {\it Five-hundred-meter Aperture Spherical Telescope (FAST)} \cite{2024ATel16505....1Z} and more than 100 bright bursts with a fluence larger than 10 Jy ms with small European dishes \cite{2024ATel16565....1O}.  The hyper-activity accelerated its localization with interferometry arrays -- an initial localization at the arcsecond level with {\it MeerKAT} \cite{2024MNRAS.533.3174T} and the subsequent precise localization at the sub-arcsecond level with the {\it European VLBI Network (EVN)} were achieved  \cite{2024ATel16542....1S}. Optical spectroscopic observations then determined that \target~is associated with a dwarf galaxy at the redshift $z = 0.13$ \cite{2024ATel16613....1B}. 

{\it CHIME} detected bright bursts in 400--800 MHz between 2024 January 21 and 25, implying the low-frequency burst rate reached about 15 per hour for bursts with a fluence larger than 1 Jy ms. {\it FAST} detected a high burst rate (500 bursts per hour) in 1000--1500 MHz for bursts with a fluence threshold of 0.015 Jy ms around 2024 March 5 \cite{2024ATel16505....1Z}. Detections of an increased burst rate at higher frequencies, in the S-band and the C-band, respectively, were reported in 2024 Mid--April and early May \cite{2024ATel16597....1H,2024ATel16620....1L}. These reports likely indicate an evolution of the high burst rate episodes, from low frequencies to high frequencies in time sequence, implying that the FRB central engine probably had been evolving rapidly.  

\begin{figure}[H]
 \centering
\includegraphics[angle=0,width=5.50in]{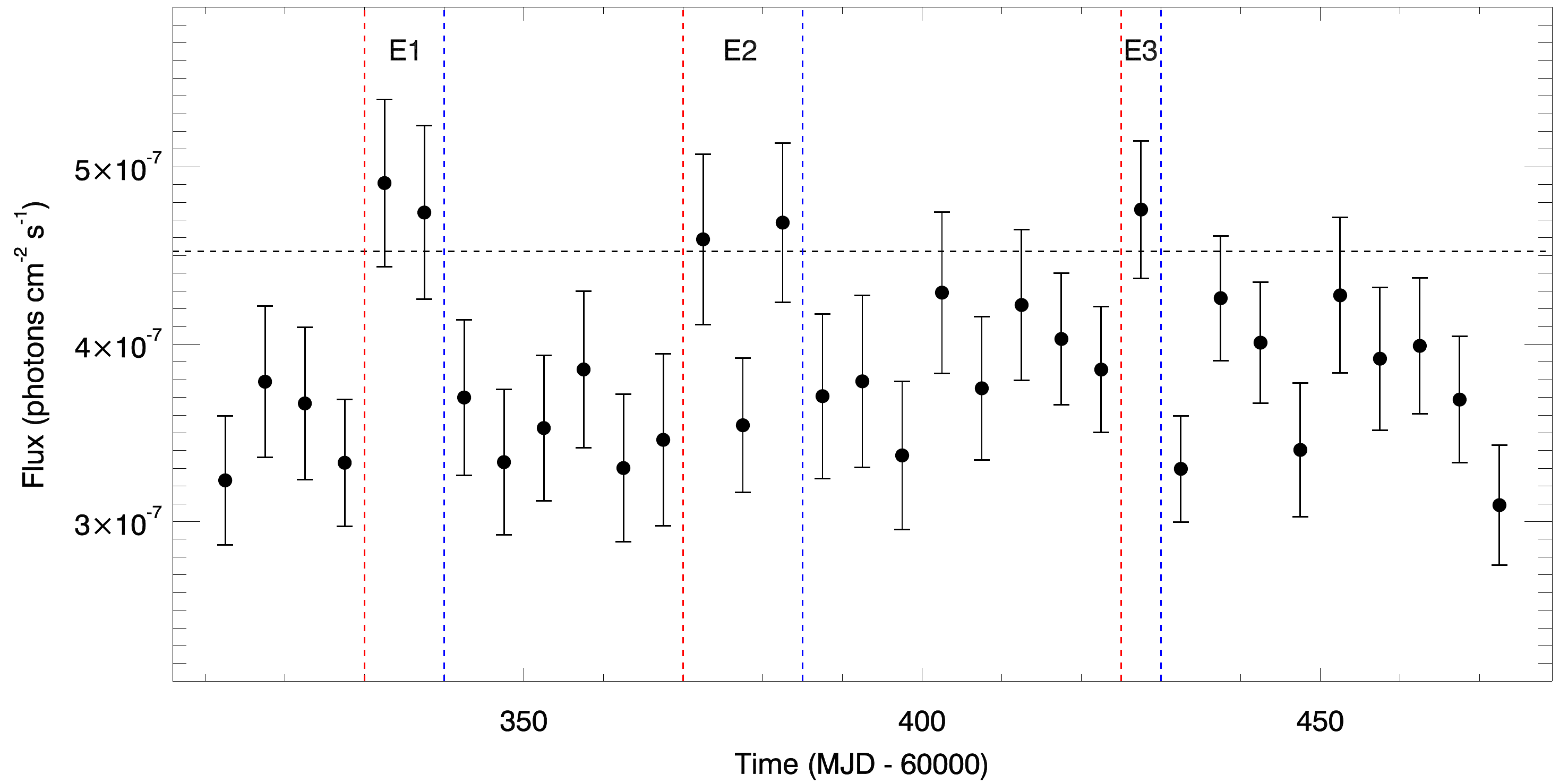}
 \caption{
 \footnotesize \textbf{Fermi/LAT \gr\ 5-day averaged light curves extracted from 0.1$-$500 GeV photons from the direction of \target.} Three episodes, namely Episode 1 (E1: MJD 60,330-60,340), Episode 2 (E2: 60,370-60,385) and Episode 3 (E3: MJD 60,425-60,430), are marked by a pair of vertical dashed lines (in red and blue). The corresponding \gr\ flux in each episode exceeds the flux level reached at most once in the entire year of 2023, which is marked as a hirizontal dashed line, indicating enhanced \gr\ emission from sources in the FRB field.}
 \label{fig:monitoring}
\end{figure}

Observations of these high burst rate episodes stimulated us to search for multi-wavelength counterparts on timescales of 5 days and longer. The search for \gr\ counterparts with the {\it Fermi} Large Area Telescope (LAT) was the top priority because of its daily regular coverage of the source and our suspect of the role of extreme relativistic boosting \cite{2023ApJ...959...89Z,2024arXiv240212084Y}. With an aperture photometric analysis of the LAT data by extracting photons in the central point-spread-function (PSF) (68\% PSF radius, dependent on photon energy) \cite{2024ATel16594....1X,2024ATel16630....1X}, we have identified three 5--15 day episodes, namely Episode 1, 2 and 3, here after named E1, E2, and E3, during which apparent enhancements of the $>$ 100 MeV \gr\ emission in the direction of \target\ were found (Figure~\ref{fig:monitoring} and Methods). These episodes appear roughly coincident with the high radio burst rate episodes observed by {\it CHIME}, {\it FAST}, and the small European dishes within a week or less. The corresponding 5-day average flux level during the three episodes had been reached only once or less in the entire year of 2023 (Methods; Figure~\ref{fig:aplc}), implying that there was probably a new, variable \gr\ source emerging in the direction of the FRB during these episodes. 

\begin{figure}[H]
 \centering
\includegraphics[width=0.44\textwidth,angle=0]{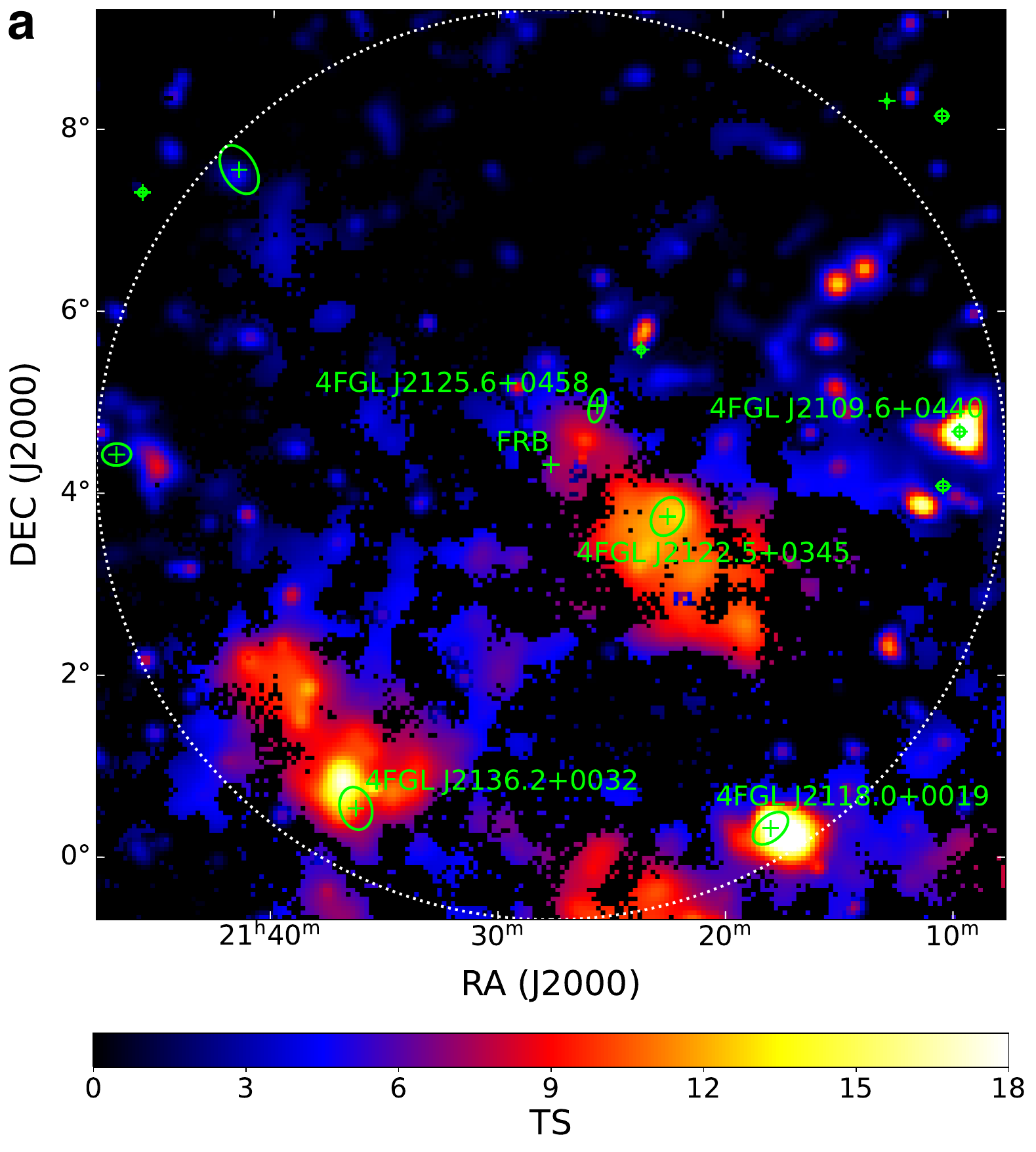}
\includegraphics[width=0.46\textwidth,angle=0]{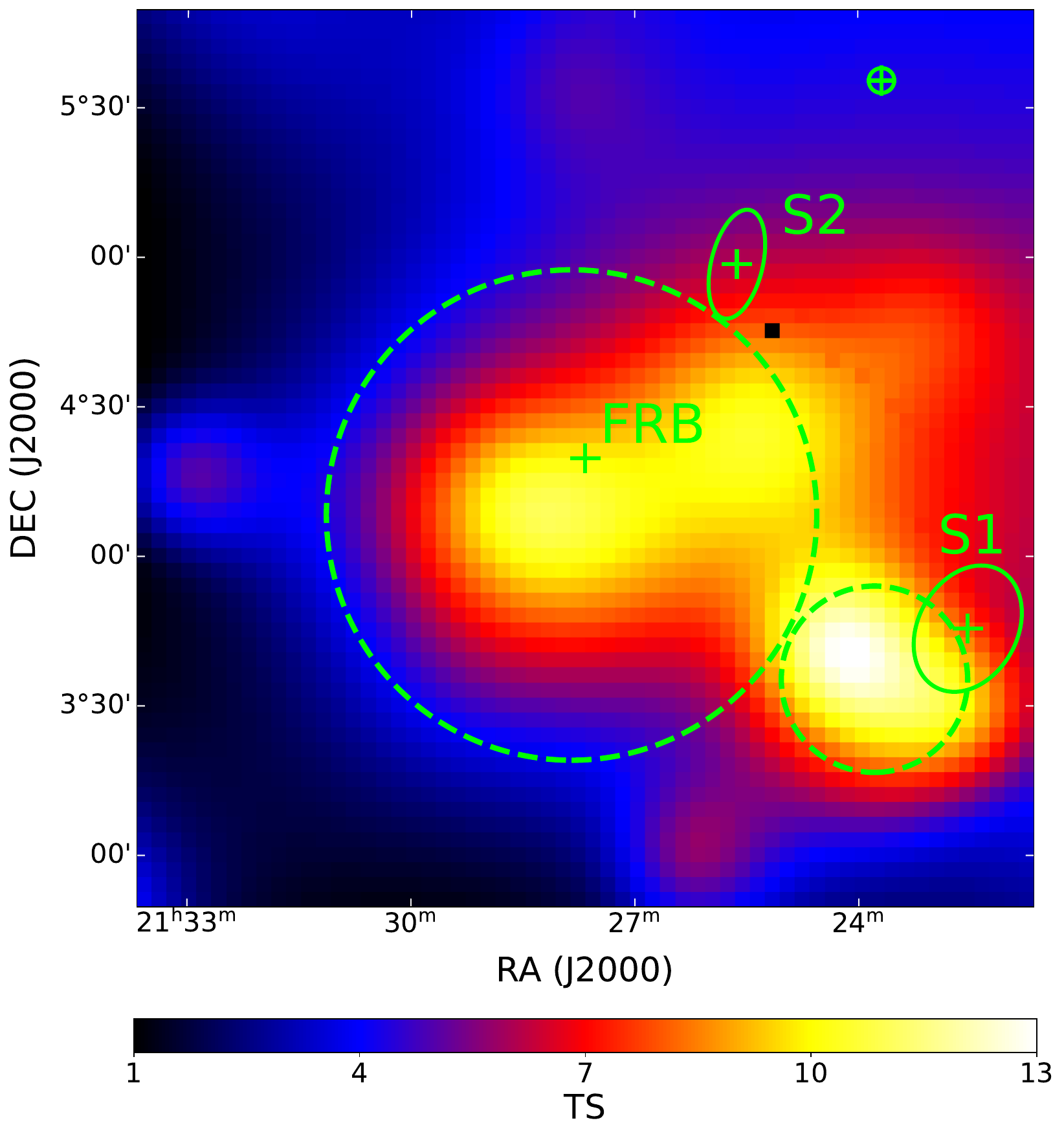}

 \caption{
 \footnotesize \textbf{The 0.1--500 GeV Fermi/LAT TS map in 2023 in the FRB FoV (Left, a) and the TS map corresponds to the third \gr\ active episode E3 (Right,b)}. The green plus sign and ellipses are the positions and the 95\% error ellipses of sources given in the Fermi source catalogue. The position of \target\ is marked as a green plus sign, while the 2$\sigma$ error circles of the possible sources which are responsible for the excessive $\gamma$-ray flux from the direction of FRB 20240114A and S1 are shown as green dashed circles in (b) centered at their best-fit positions.
 The excessive \gr\ flux at the center of the FoV is statistically consistent with the position of the FRB well within 1$\sigma$ range, against its association with S1 or S2, which are more than 2$\sigma$ away. }
 \label{fig:ts_episode}
\end{figure}

In the Fermi/LAT field-of-View of \target, based on maximum likelihood analysis which considers all 0.1--500 GeV photons in the 20$^{\circ}$ $\times$ 20$^{\circ}$ square region centered at the position of \target, only one of the Fermi sources, 4FGL J2122.5$+$0345, the second nearest known Fermi source, here after named as Source 1 (S1), has been detected during the period in 2024 when \target\ is active (Figure~\ref{fig:ts_episode}). The most nearby 4FGL catalogue source 4FGL J2125.6$+$0458, here after Source 2 (S2), was not detected with the Fermi/LAT data accumulated in 2023 and 2024, respectively. Since the LAT photon events collected in the past decades can be used to statistically rule out any of the known sources that has short-term variability, we therefore focused on an unbiased photon selection among S1, S2, and the FRB position (see Method) in order to exclude the two Fermi sources (S1 and S2) as the candidates responsible for the enhanced \gr\ emission. 

The photon counts extracted in the central PSF surrounding the FRB, on average, has $\sim$ 25 counts per day in 2024. An interval with a duration of 5 to 15 days corresponds to typically 125 to 375 photons in total falling in the central PSF. It is possible that \gr\ photons may arrive in clusters in the form of \gr\ flares. We therefore looked into photon arriving times for photon clusters in E1, E2, and E3 to identify potential short-term \gr\ flares that might produce the \gr\ enhancement. Selection of photon clusters with short arriving time intervals enforces a filtering of photons -- the selection of photons from only sources with short-term variability and stable, bright sources. At the average daily photon rate of 25--30 counts per day, it also helps remove photons from stable sources in the field-of-view if we target at cluster time scales of seconds. 

We started our investigation with photon doublets. We first noticed three pairs of double photon events (doublets) with an interval of less than 2.0 seconds in E1 and E2 \cite{2024ATel16594....1X}, which favor an association with \target~instead of an association with either of the two nearby Fermi catalogues sources \cite{2023arXiv230712546B} (see Methods; Figure \ref{fig:2p}). For a pure Poisson fluctuation of photon counts in the direction of \target, each doublet arriving in less than 2.0 seconds corresponds to a significance around 5$\sigma$ \cite{1983ApJ...272..317L,2021NatAs...5..385F} (see Methods). Actual photon series in observations won't be pure Poisson since there are sources in the field-of-view. An investigation of the actual photon events collected in the mission-long data gives the estimate of the chance of observing such a short-duration doublet as less than one out of 500 photons; we expect less than one pair of doublet ($<$ 2 seconds) in any interval of 10 days. The three episodes of apparent enhanced \gr\ emission and the detection of doublets with an interval $<$ 2 seconds therefore hint an underlying variable \gr\ source that can potentially generate flares on seconds time-scale. 

We continued the same investigation when we identified Episode 3. We found a remarkable photon quadruplet, which arrived in an interval of only 15.6 seconds for photons centered specifically at the Equal-Angular-Distance (EAD) position to the FRB, S1 and S2, namely Position C as defined (see Methods). This quadruplet was initially found as a GeV triplet from photons centered at the FRB position. An additional photon was identified when a slightly larger sky region is considered. The quadruplet was led by a 1.01 GeV photon only 0.3\arcdeg\ away from the precise FRB position, followed by three 115--141 MeV photons arrived roughly 11--16 seconds later (as a 5-s triplet alone) \cite{2024ATel16630....1X}. The significance of such a quadruplet with a duration like this as a timing signal is 6.1$\sigma$ for pure Poisson fluctuations \cite{1983ApJ...272..317L,2021NatAs...5..385F}. 

\begin{figure}[H]
 \centering
\includegraphics[angle=0,width=0.450\textwidth]{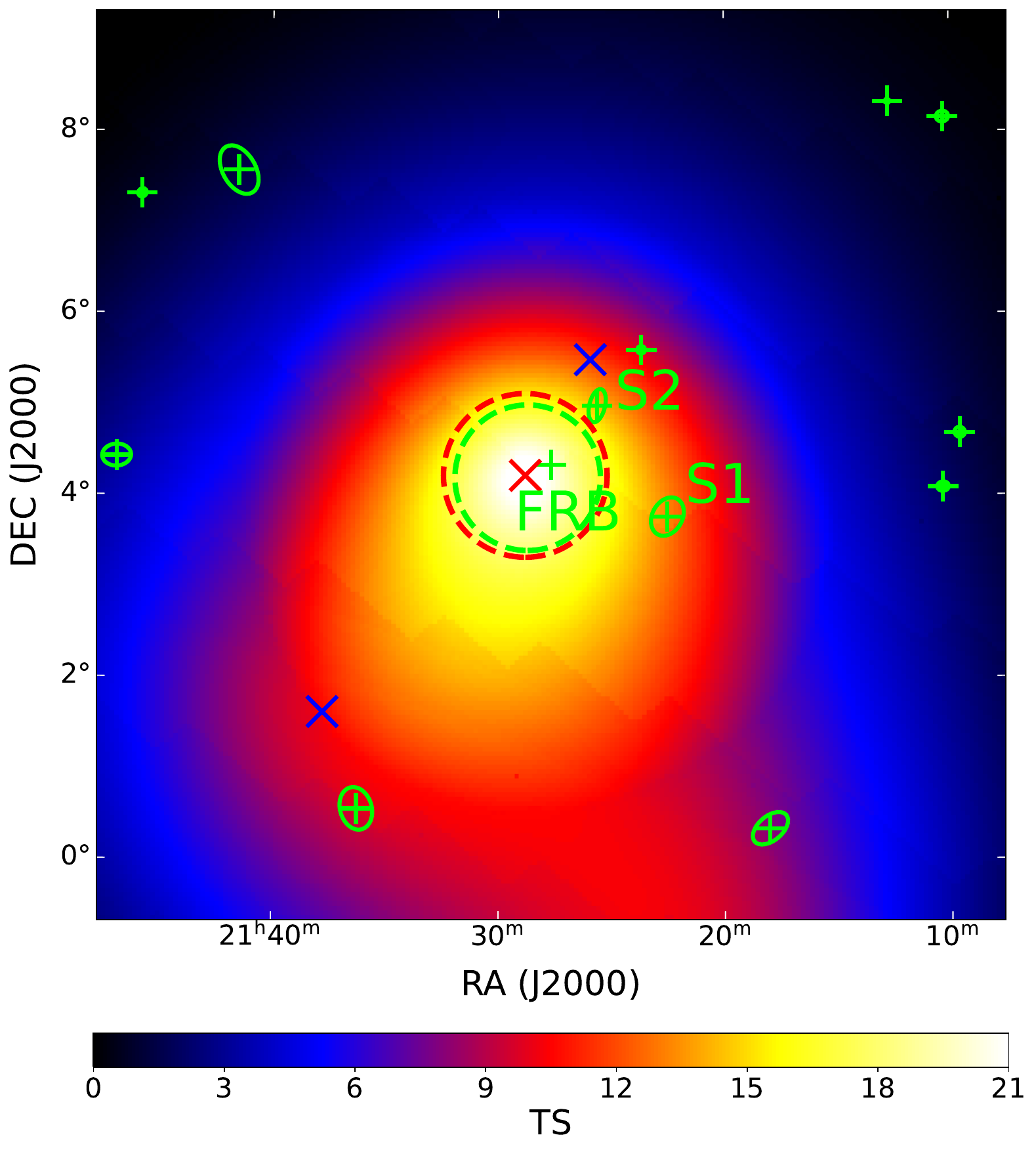}
\includegraphics[angle=0,width=0.51\textwidth]{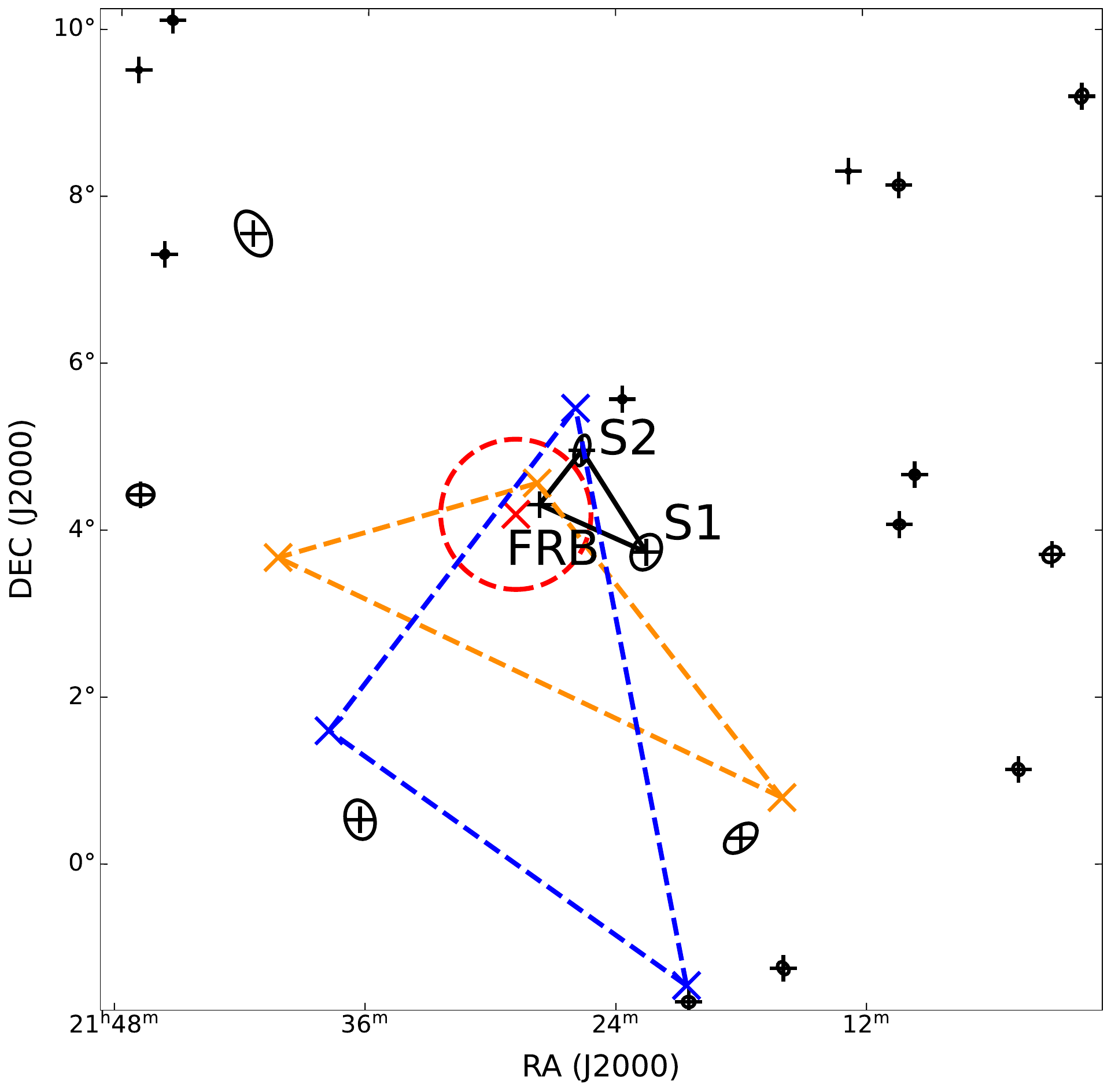}

 \caption{
 \footnotesize \textbf{The Fermi/LAT TS map of the FRB FoV during the 15.6 s GeV \gr\ flare (Left,a) and the positions of the photons of the GeV quadruplet and the 29-s triplet (Right,b).} The position of the GeV photon is marked as a red cross and its 68\% point-spread-function circle is marked as a circle in red dashed line in (a). The position of the GeV photon and its 68\% range are marked as a red cross and circle, the positions of the three 115-141 MeV photons in the quadruplet are shown as a blue triangle, and the positions of an additional 29-s triplet are shown as a yellow triangle in (b). Fermi sources S1 and S2 are outside the circles or the triangles, indicating they are unlikely the source producing the 15.6-s quadruplet and the 29-s triplet.}
 \label{fig:ts_flare}
\end{figure}

We then performed the maximum likelihood imaging analysis of the Fermi/LAT data during the E3 and the quadruplet interval (see Methods). The analysis corresponding to the quadruplet interval yielded the position of the \gr\ flare at R.A.=322.2\arcdeg, Decl.=4.2\arcdeg\, with a 1$\sigma$ nominal uncertainty of 0.5\arcdeg\ and a 2$\sigma$ nominal uncertainty of 0.8\arcdeg\  (Figure~\ref{fig:ts_flare}). The best-fit position is 0.3\arcdeg\ away from the FRB position. The maximum likelihood analysis yields a TS value of $\sim$ 21 (see Methods). The nearby Fermi catalogued source S2 is 1.1\arcdeg away and S1 is 1.6\arcdeg\ away, suggesting that the GeV \gr\ flare unlikely came from known sources in the Fermi/LAT catalogue within the field-of-view. Furthermore, a GeV \gr\ quadruplet this short has never been observed previously in the Fermi/LAT's mission-long events collected in the same angular range in the past, despite the detection of a historical GRB 20141028A which is 5.2\arcdeg\ away from the EAD position. The non-detection of any previous quadruplet this short other than the detection of the previous GRB 20141028A indicate the 15.6-s quadruplet is a highly significant gamma-ray flaring signal with a peak flux  comparable to a weak GRB. The association of the GeV photon and the following MeV triplet are also highly significant, since we found the GeV photons from the FRB direction have been rare in the past; the GeV quadruplet is the shortest seen in the mission-long data (see Methods). Our analysis therefore implies that the GeV \gr\ flare is probably from a new, variable source consistent with the FRB position. Independent evidence for such a bright variable source is also collected by the maximum likelihood analysis of the LAT events during the E3 (Figure~\ref{fig:ts_episode}) and the detection of an independent 29-s triplet, which also favours an association with the FRB (see Figure~\ref{fig:ts_flare} and Methods). 

To date, multi-wavelength counterparts other than the radio wavelength have never been detected in FRBs of cosmological distances \cite{ZhangARNPS24,2017ApJ...846...80S,zhangzhang17,2019ApJ...885...55S,2023A&A...675A..99P}. Through a strategy starting with the identification of \gr\ active episodes, we have found evidence of coincident \gr\ flares associated with the hyper-active \target. The best known variable extragalactic \gr\ sources include \gr\ blazars \cite{2017Natur.552..374R,2019ApJ...877...39M,2022PASP..134j4101W} with hour-to-day variability timescales, and \gr\ bursts (GRBs) or magnetar giant flares with hourly or shorter variability timescales \cite{2013ApJS..209...11A,2019ApJ...878...52A,2021NatAs...5..385F}. At $z=0.13$, the isotropic \gr\ luminosity of the quadruplet-defined 15.6-s GeV \gr\ flare is $\sim 10^{48} \ {\rm erg \ s^{-1}}$. Such a high-luminosity has been only reached by GRBs and extremely luminous active galactic nuclei known as blazars. However, the shortest variability timescale of previously detected blazars was $\sim$1 min \cite{2019ApJ...877...39M,2022A&A...668A.152P}, much longer than the seconds timescale of the doublets and triplets identified in Episodes 1, 2 and 3. 
Also, these short flaring activities and gamma-ray enhancement episodes span in time for at least three months, suggesting that this high-luminosity \gr\ source must remain active over such a period of time. This is different from GRBs whose afterglow emission is known to decay rapidly. The peculiar \gr\ flare(s) detected from the direction of \target\ therefore suggests either a completely new type of source or a very rare type of known source, whose activities are somewhat related to the activation of \target.


The existence of such a \gr\ flaring source in association with the FRB source is inconsistent with the known engine that can potentially power FRBs, namely, magnetars with a rotation period of seconds that powered the Galactic FRB with a lower radio luminosity \cite{CHIME20,STARE2-20}. Our finding implies two possibilities (see Methods). First, the \gr\ flaring source may be the source of \target. However, such a counterpart is very rare for an earth observer because none of the other FRB sources were observed to be associated with \gr\ emission \cite{zhangzhang17,yang19}. One possible scenario is that the source is a new-born millisecond pulsar. Such a pulsar with a spin-down luminosity below $1.5\times 10^{45} \ {\rm erg \ s^{-1}}$ but occasionally making flares with luminosities of $\sim 10^{48} \ {\rm erg \ s^{-1}}$
may explain the observations, even though certain model constraints are needed. Another scenario is a new-born microblazar with an extremely high \gr\ luminosity that also emits FRBs. Such a scenario has not been theoretically explored before, but is not impossible (Methods). Alternatively, the FRB engine may be similar to most other FRB engines but it is located close to a previously-dormant, recently-activated or even a new-born \gr\ source (e.g. a blazar). The energetic \gr\ flux and its associated outflow may indirectly trigger the engine to make repeated FRB bursts \cite{zhang17}. Further observations of \target, its \gr\ counterpart, and its associated compact radio continuum counterpart \cite{2024ATel16695....1Z}, will help differentiate among these possibilities.


\bigskip
\bigskip
\bigskip

\begin{addendum}
\item This work made use of data from the \textit{FERMI} telescope, which is managed by NASA Goddard Space Flight Center. The Large Area Telescope (LAT) is the main instrument onboard \textit{FERMI}, we thank the Fermi Science Support Center (FSSC) for providing the public LAT data\footnote{\footnotesize https://fermi.gsfc.nasa.gov/cgi-bin/ssc/LAT/LATDataQuery.cgi}. W.Y. would like to acknowledge the support from the Natural Science Foundation of China under grant No.12373050. 

 \item[Author contributions] 
Y. X. analyzed the Fermi/LAT raw data; both Y. X. and W.Y. analyzed the Fermi/LAT photon events; W.Y. proposed the project and the search strategy and methods; Z. Y. and X. Z. investigated multi-wavelength data; B. Z. conducted theoretical calculations and led the interpretations; W. Y. contributed partly to the interpretations. W. Y., Y. X., and B. Z. contributed to the writing of the first, primary draft; all authors contributed to the writing of the draft and the  discussions of the results. 
 
 \item[Competing interests] The authors declare that they have no competing financial interests.

\item[Correspondence] Correspondence and requests for materials should be addressed to W. Yu (wenfei@shao.ac.cn) and B. Zhang (zhang@physics.unlv.edu).

\item[Data availability]
FERMI/LAT data are directly available from the FERMI data archive in the public domain. This research has made use of \emph{FERMI} data, which are publicly available and can be obtained through the High Energy Astrophysics Science Archive Research Center (HEASARC) website at \url{https://heasarc.gsfc.nasa.gov/W3Browse/}

\item[Code availability]
Much analysis for this paper has been undertaken with publicly available codes and the details required to reproduce the analysis are contained within the manuscript.

\end{addendum}

\clearpage

\newpage 

\begin{methods}


\section{Fermi/LAT Gamma-ray observations}
The {\it Fermi} satellite has an orbital period of 96.5 minutes. The main instrument, the Large Area Telescope (LAT), is a pair conversion telescope designed to cover the energy band from about 20 MeV to greater than 300 GeV. It continuously scan the whole sky every two orbits. In the first 14 years of operation, Fermi/LAT survey observations had yielded source detections, which is listed as the updated Fermi catalogue 4FGL-DR4. We investigated those catalogue sources nearby \target. We have found 1 source within 1\arcdeg, 2  additional sources within 2\arcdeg, and 6 additional sources within about 4\arcdeg\--5\arcdeg. Our search for potential \gr\ emission from the newly-discovered \target\ requires the understanding of the \gr\ emission of these known \gr\ sources during the period of the activation of \target. 


We selected relevant LAT photon events from the updated \textit{Fermi} LAT Pass 8 database (P8R3). 
Those 0.1--500 GeV events within a circular region centered at \target\ with a radius
of 15$^{\circ}$ were downloaded from the LAT data server. The start and end time of the time window regarding this report are 2008-08-04 15:43:39 (UTC) and 2024-06-14 00:00:00 (UTC), respectively. Our monitoring of the source through regular Fermi/LAT photometry analysis has been kept going. We included events in the \texttt{SOURCE} event class with zenith angles less than 90 degrees to avoid the Earth's limb contamination, and excluded events with `bad' quality flags. These selection criteria are recommended by the LAT team\footnote{\footnotesize http://fermi.gsfc.nasa.gov/ssc/data/analysis/scitools/}. The selection yielded the full 0.1 -- 500 GeV LAT photon event sample in our analysis. 

\section{Fermi/LAT Data Reduction}
\subsection{Fermi/LAT Photometry Analysis and Results\\}
We first performed an aperture photometry analysis to the LAT data to gain a quick search for \gr\ emission potentially associated with the bursting activities of \target. Since the aperture photometry analysis is a background-model independent analysis, which do not consider possible variability in the source background, we only used LAT observations since 2023 in our photometry analysis. We selected the 0.1--500 GeV LAT photons within the energy-dependent Point Spread Function\footnote{\footnotesize https://www.slac.stanford.edu/exp/glast/groups/canda/lat\_Performance\_files/PSF\_Ave68Energy\_P8R3\_SOURCE\_V3Total.txt} (PSF) after 2023-01-01 00:00:00 (UTC) (MJD 59945) and binned them into 5-day time intervals. The exposures of the circular sky region with a typical radius of 5$^{\circ}$ (the widest PSF is 5.3 degrees for photons with energies at 100 MeV) determine the photon counts in each of the bins. We assumed a power-law source spectrum with an index of 2.1, and the source fluxes were therefore obtained by dividing the photon counts by the exposures of the corresponding time bins obtained with \textit{gtexposure} in the {\tt Fermitools}. The above routine analysis resulted an aperture photometry light curve, as well as the histogram of the 5-day averaged fluxes during 2023, which are shown in Figure~\ref{fig:aplc}. Based on the 2023 5-day averaged light curve, we determine the maximum 5-day averaged flux as the threshold, to determine if there is enhanced emission from the direction of \target\ in 2024; when any 5-day averaged flux exceeds the flux threshold which reached once in 2023, we then identify the corresponding 5-day interval in 2024 as an episode of enhanced \gr\ emission from the FRB direction. 

With the 2024 LAT data, we have identified three active episodes of enhance \gr\ flux in 2024 \cite{2024ATel16594....1X,2024ATel16630....1X}, corresponding to MJD 60,330--60,340, MJD 60,370--60,385, and MJD 60,425--60,430, respectively. These episodes are named Episode 1 (E1), Episode 2 (E2), and Episode 3 (E3), respectively. They are shown by vertical dashed lines in Figure~\ref{fig:monitoring}. During these three active episodes, the \gr\ flux as seen from the aperture photometry light curves are approximately equal or above the highest 5-day averaged flux detected in 2023, suggesting enhanced \gr\ flux from the FRB direction probably during at least part of the three episodes in 2024. We then investigated the photon series in the episodes for evidence of short-term variability that might accompany the radio activity of the FRB source. 

\subsection{Maximum Likelihood Analysis of Fermi/LAT data \\}
We performed the standard maximum likelihood analysis of the LAT data independent of the above photometry analysis to find active \gr\ sources during the three episodes (E1, E2, and E3). In addition, we also performed the same maximum likelihood analysis on the 2023 data to identify bright sources in the recent past in the FRB field-of-view (see Figure~\ref{fig:ts_episode}) as well as the E3 data, including the data corresponding to the 15.6-s GeV flare interval to locate the source responsible for the short flare (see Figure~\ref{fig:ts_episode} and Figure~\ref{fig:ts_flare}) as compared with known nearby Fermi catalogue sources. 
It is worth noting that all the maximum likelihood analysis of Fermi/LAT data are not biased to any specific position (no positional selection effect). This also applies to the maximum likelihood analysis for the data during the 15.6-s GeV flare and the entire E3, which included those outside the central PSF.

In the maximum likelihood analysis, we included all the 4FGL-DR4 sources \cite{2023arXiv230712546B} within 20 degrees from \target\ in the source model. We adopted those measured positions and the spectral parameters of these sources in the catalogue. The spectral model file {\it gll\_iem\_v07.fit} and the spectral file 
{\it iso\_P8R3\_SOURCE\_V3\_v1.txt} were used in our analysis, in order to account for the contributions of the Galactic and extragalactic diffuse emission, respectively. The normalizations of the two diffuse components were always set as free parameters in our analysis.

The Region of Interest (RoI) was defined as a 20$^{\circ}$ $\times$ 20$^{\circ}$ square region centered at the position of \target. We first calculated the 0.1--500 GeV Test Statistic (TS) map of a $\mathrm{10^{o}\times10^{o}}$ region centered at the position of \target\ during the entire 2023 (MJD 59,945--60,310). All sources in the original source model, except those within $\mathrm{5\arcdeg}$ from the FRB, were considered in our source model and subsequently removed from the TS map. The resulted TS map is shown in the left panel of Figure~\ref{fig:ts_episode}. In such a maximum likelihood analysis of LAT data, the TS value is a measurement of the fit improvement for including a source at its position, which approximately corresponds to the square of the detection significance \cite{2020ApJS..247...33A}. There are four catalogue sources, namely J2122.5+0345 (S1), J2109.6+0440, J2136.2+0032, and J2118.0+0019, are significantly revealed in the map, while the most nearby catalogue source J2125.6+0458 (S2) is not significantly revealed in the 2023 TS map. S1 had a 0.1--500 GeV photon flux of 6.2 $\times$ 10$^{-9}$ photons cm$^{-2}$ s$^{-1}$ and a TS value of 6. S2 has a TS value of 0, and the 95\% photon flux upper limit for the 0.1--500 GeV \gr\ emission of S2 is 1.1 $\times$ 10$^{-9}$ photons cm$^{-2}$ s$^{-1}$. No significant \gr\ emission was detected around the FRB position with a TS value above 1.0. The 95\% photon flux upper limit for the 0.1--500 GeV \gr\ emission at the FRB position is 8.7 $\times$ 10$^{-9}$ photons cm$^{-2}$ s$^{-1}$, corresponding to an energy flux upper limit of 3.4 $\times$ 10$^{-12}$ ergs cm$^{-2}$ s$^{-1}$. Our analysis of the LAT data accumulated in 2024 indicates that S1 and S2 are not significantly revealed. The 95\% photon flux upper limit for the 0.1--500 GeV \gr\ emission from S1 and S2 is 1.0 $\times$ 10$^{-8}$ photons cm$^{-2}$ s$^{-1}$ and 2.4 $\times$ 10$^{-9}$ photons cm$^{-2}$ s$^{-1}$, with a TS value of 1 and 0, respectively. The \gr\ emission around the FRB position is not significantly detected in the 2024 data, with a TS value of 0. The 95\% photon flux upper limit is obtained as 5.1 $\times$ 10$^{-9}$ photons cm$^{-2}$ s$^{-1}$, corresponding to an energy flux upper limit of 2.0 $\times$ 10$^{-12}$ ergs cm$^{-2}$ s$^{-1}$. 

We then calculated the 0.1--500 GeV TS maps of a $\mathrm{3^{o}\times3^{o}}$ region centered at the FRB position during E1, E2, and E3, to identify underlying variable \gr\ source(s). A marginal detection of an excess \gr\ emission can be observed around the FRB position during E3, with a TS value of $\sim$9 (see the right panel of Figure~\ref{fig:ts_episode}). In addition, \gr\ emission is also visible around the position of S1. With these detections in E3 in mind, to determine potential \gr\ flares from the FRB later on, we need to exclude S1 as the origin. S2 is not the source of the enhanced \gr\ emission seen in photometry measurements during E3. 

We then ran \textit{gtfindsrc} in the {\tt Fermitools} to the 0.1--500 GeV full event sample in E3 to update the position of the catalogue source S1 in the current data set, and obtained R.A.=320.9\arcdeg, Decl.=3.6\arcdeg\ (equinox J2000.0) for S1, with a 1$\sigma$ nominal uncertainty of 0.2\arcdeg. We also ran \textit{gtfindsrc} to determine the position of the enhanced \gr\ emission revealed around the FRB position, and obtained R.A.=322.0\arcdeg, Decl.=4.1\arcdeg, with a 1$\sigma$ nominal uncertainty of 0.5\arcdeg. The 2$\sigma$ error circles of these two fitted positions are plotted as dashed green circles in the right panel of Figure~\ref{fig:ts_episode}. The FRB position is 0.2\arcdeg away from the fitted position, well within the 1$\sigma$ error circle to the source producing the enhanced \gr\ emission, and both S1 and S2 are 2$\sigma$ away from the derived position. 

We then re-performed the likelihood analysis to the LAT data during E3 using the updated source model, which included the source producing the excess emission around the FRB and the source S1 considered as point sources at their best-fit positions. The spectral normalizations of the catalogue sources within 5\arcdeg\ from the FRB were let free in the analysis, while all the other parameters were fixed at their catalogue values. The \gr\ emission of S1 is described with a log-parabola model in the \fermi\ catalogue. We fixed its spectral shape parameters and only let the spectral normalization free. For the FRB counterpart candidate, we adopted a simple power-law model to describe its \gr\ emission, and the power-law index $\Gamma$ and the normalization were set free. We obtained 0.1--500 GeV spectral index $\Gamma$ of 2.7$\pm$0.7 and a photon flux of 6.1$\pm$4.4 $\times$ 10$^{-8}$ photons cm$^{-2}$ s$^{-1}$ for the candidate, with a TS value of 5. The photon flux of the candidate is higher than that of S1, which is measured as 4.0$\pm$2.7 $\times$ 10$^{-8}$ photons cm$^{-2}$ s$^{-1}$ with a TS value of 4. Our analysis supports the emergence of a new \gr\ source in the direction of the FRB during E3. If the candidate source is indeed the FRB source at z=0.13 \cite{2024ATel16613....1B} with a luminosity distance of $\sim$610 Mpc, then the 5-day average luminosity is about (1.1$\pm$0.7) $\times$ 10$^{45}$ ergs s$^{-1}$, as given by the maximum likelihood analysis.

\subsection{Photon Series: Doublets, Triplets and Quadruplets\\}
We investigated double photon events (doublet), triple photon events (triplet), and quadruple photon events (quadruplet) in each of the three episodes. Preliminary results have been reported \cite{2024ATel16594....1X,2024ATel16630....1X}. Initially, we only focused on doublet and triplet, until we identified the quadruplet when investigated a triplet. The time intervals of sequenced, overlapping doublet, triplet, and quadruplet can be obtained straightforwardly. We applied the same estimation of the significance described in \cite{2021NatAs...5..385F,2023A&A...675A..99P} to the LAT data to identify \gr\ photon clusters which arrived within short time intervals during the episodes E1, E2, and E3, which is based on the Li-Ma formula \cite{1983ApJ...272..317L}. The 0.1--500 GeV LAT events within the energy-dependent 68\% PSF were selected as in the photometry analysis. We computed the time intervals $\mathrm{\Delta}t_{i}$ for each doublet and each triplet, formed by two and three consecutive events as,
\begin{equation}
\mathrm{\Delta}t_{i,2p} = t_{i+1}-t_{i}
\end{equation}
\begin{equation} 
\mathrm{\Delta}t_{i,3p} = t_{i+2} - t_{i}
\end{equation}
respectively. These estimates are based on pure Poisson fluctuation (see Figure~\ref{fig:li_ma} in Methods). They tell us about the interval threshold which defines a significant doublet, triplet and quadruplet, respectively. 

We first found two independent doublets during E1 and E2, with $\mathrm{\Delta}t_{2p}$ of 0.7 and 1.8 seconds, respectively. On MJD 60369, there exists another photon doublet of a similar interval with $\mathrm{\Delta}t_{2p}$ of 1.6 seconds. We also include it in our investigation, as these 5-day windows should be regarded as a rough description. The positions of these photons favor an association with \target\ rather than an association with S1 or S2. The sky positions of the six LAT photons are plotted in Figure~\ref{fig:2p}, with their parameters listed in Table~\ref{tab:2p}. The photon with the maximum energy among them is 278 MeV, and the corresponding 68\% PSF radius is $<$3.4$^{\circ}$. Constraints from these photon doublets are limited by rather large radii of the PSF at 100-200 MeV photon energies. 

In E3 we found two clusters of triplets with $\mathrm{\Delta}t_{3p}$ of only 15.6 and 29.0 seconds, respectively. The parameters of the photons are listed in Table~\ref{tab:3p}, and the positions of these photons are plotted in Figure~\ref{fig:ts_flare}. The photon triplet with $\mathrm{\Delta}t_{3p}$ of 15.6 seconds was led by a GeV photon. Its position is marked as the red plus sign in Figure~\ref{fig:ts_flare}, which is only 0.3\arcdeg\ away from the FRB. The 68\% PSF radius for the GeV photon energy is about 0.9\arcdeg. This argues for the association of the GeV photon with the FRB, and consequently the association of the GeV triplet/quadruplet(see below) with the FRB as well. We also found that the positions of the triplet photons are also in favor of an association with the FRB rather than with S1 or S2, as the center of the triangles formed by triplet photon positions are all on the FRB side (Figure~\ref{fig:ts_flare}).  

By analyzing the LAT data within the 68\% PSF during the nearly 16 years in the past, we can estimate the chance of finding such photon doublets or triplets with $\mathrm{\Delta}t$ less than $\mathrm{\Delta}T$ by accident in the real data. From the actual photon events extracted from the direction of the FRB, we found that those photon doublets with $\mathrm{\Delta}t_{2p}$ less than 2 seconds contribute to only 2.1$\times$10$^{-3}$ of the total number of photons collected, and the triplets with $\mathrm{\Delta}t_{3p}$ less than 16 seconds contribute to 1.7$\times$10$^{-4}$ of the total number of photons. In addition, we also found the GeV photon triplet with $\mathrm{\Delta}t_{3p}$ of 15.6 seconds is the only GeV triplet with $\mathrm{\Delta}t_{3p}$ of $<$16 seconds in the entire photon sample extracted from the FRB direction. Even without a detailed consideration of that these doublets and triplets were clustered in the 5-15 day window as E1, E2, and E3 in early 2024, our investigation suggests that these doublets and triplets we found indicate flaring signals from an underlying source in the FRB FoV that was not active in the past, against an association with previously existing sources S1 and S2. 

The two nearby \fermi\ catalogue sources S1 and S2 are 1.4\arcdeg and 0.8\arcdeg\ from the position of \target, respectively, which may contribute to contaminated \gr\ emission at the FRB position. To perform a photon selection unbiased to the three sources, namely the FRB, S1 and S2, we extracted photons centered at the specific position C with an Equal Angular Distance (EAD) of $\rm r$ from each of the FRB, the source S1, and the source S2, with the $\rm r$ being 0.74\arcdeg (see Figure~\ref{fig:cpoint}). We selected the 0.1--500 GeV LAT events within PSF$+$0.74\arcdeg\ and then searched for clustered \gr\ photons within short time intervals during the three flaring episodes. In this new photon series from a larger FoV, we found that there is a fourth 137.7 MeV photon arrived right in the 15.6-s interval of the triplet we identified previously (see Table~\ref{tab:3p}, the photon has $\mathrm{\Delta}t$ of 10.6 seconds to the leading GeV photon), joining the triplet to form a quadruplet instead with $\mathrm{\Delta}t_{4p}$ ($t_{i+3} - t_{i}$) of only 15.6 seconds. By considering the total LAT event data within PSF$+$0.74\arcdeg\ from the C point, the percentage of observing such a photon quadruplet is only 3.3$\times$10$^{-5}$. Furthermore, we investigated photons collected from the FRB direction surrounding Point C in the entire 16 years' LAT observations; we found a total of five quadruplets with $\mathrm{\Delta}t_{4p}$ less than 16 seconds, while the other four quadruplets overlapped, corresponding to a historical \gr\ burst GRB 141028A. It is worth noting that even in the mission-long photon list extracted in the larger region surrounding Point C, the GeV quadruplet observed during E3 is still the only quadruplet with a GeV photon in less than 16 seconds, indicating the peculiarity of the GeV flare is unprecedented. 

We further calculated the 0.1--500 GeV TS map with the LAT data during 15.6-s quadruple flare, with a source model including only the two Galactic and extra-galactic diffuse background emission models. The resulting TS map is shown in the left panel of Figure~\ref{fig:ts_flare}. The excess \gr\ emission is revealed in the map with a TS value of $\sim$21. The best-fit position of the excess emission during the 15.6 seconds has coordinates of R.A.=322.2\arcdeg, Decl.=4.2\arcdeg\, with a 1$\sigma$ nominal uncertainty of 0.5\arcdeg\ (the 2$\sigma$ error circle is plotted as a green dashed circle in Figure~\ref{fig:ts_flare}). The FRB is 0.3\arcdeg\ away from the best-fit position and within the 1$\sigma$ nominal uncertainty. We noticed that the Fermi LAT likelihood analysis tool can not yield a flux measurement as small as 15.6 seconds. The position is also consistent within uncertainties with the best-fit position for the enhanced \gr\ emission during E3. The FRB is 0.2\arcdeg\ away from the best-fit position obtained during E3, also within uncertainties (see the green dashed circle in the right panel of Figure~\ref{fig:ts_episode}).

To measure the flux of the 15.6-s GeV flare, we performed the likelihood analysis to the LAT data during the 100-s time bin which covers the 15.6-s interval of the quadruplet, with a source model including only the two Galactic and extragalactic diffuse background \gr\ emission models and a point source at the fitted position obtained during the 15.6-s interval of the photon quadruplet. The normalizations of the two diffuse components were also set free. A spectral index $\Gamma$ of 2.4$\pm$0.7 and a 0.1--500 GeV photon flux of 8.3$\pm$5.7 $\times$ 10$^{-6}$ photons cm$^{-2}$ s$^{-1}$ were obtained for the gamma-ray candidate of the FRB, with a TS value of 12. This spectral result corresponds to a 0.1--500 GeV energy flux of 5$\pm$5 $\times$ 10$^{-9}$ ergs cm$^{-2}$ s$^{-1}$. At a redshift of z=0.13 \cite{2024ATel16613....1B} (a luminosity distance of $\sim$610 Mpc), the luminosity is about 2.1 $\times$ 10$^{47}$ ergs s$^{-1}$. The number of photons collected in the central PSF at the FRB position is 151 in E3. If such 15.6-s \gr\ flares are rare, we would not have detected one of them when the number of photons received is on the order of 30 per day during a daily exposure time in the range between 9 and 10 hours during E3. Therefore, it is likely the source emit plenty flares. If we attribute all the 100-s flux to the 15.6-s GeV flare alone, then the luminosity of the GeV flare is 1.3$\times$ 10$^{48}$ ergs s$^{-1}$. Refer to the 5-day average luminosity obtained in E3 of about 1.1 $\times$ 10$^{45}$ ergs s$^{-1}$, the source need to produce 366 such 15.6-s GeV flares during the 5-day episode. So the inferred daily \gr\ flare rate is likely on the order of 100 per day or about 4 per hour, in comparison to the radio burst rate of hundreds per hour \cite{2024ATel16505....1Z}. This will have an impact on the future strategy for searching high energy burst counterparts of those radio bursts.

\subsection{Significance of Temporal and Spatial Coincident Associations\\}

We have shown that the three \gr\ enhancement episodes appeared to associate with those reported high burst rate episodes on time scales of 5–15 days. Such high \gr\ emission episodes on timescales of 5 days or so have only been observed once in 2023. On short time scales, we have found a GeV quadruplet, a triplet, and a few doublets with intervals below 30 seconds. Such short \gr\ flares, have never been seen in gamma-ray sources such as blazars, although we might have been seen before in GRBs, but it requires a GeV GRB. In addition, such a short GeV flare as the quadruplet has never been observed in history from known Fermi/LAT sources in the field-of-view. The temporal coincidence of \gr\ activities with the activation of \target\ can not be simply explained. 

The significance of short \gr\ flares as a timing signal from a source can be addressed by the Likelihood Ratio (LR) method\cite{1983ApJ...272..317L}. We have applied the same Likelihood Ratio (LR) method described in \cite{2021NatAs...5..385F}, which is based on the Li-Ma formula \cite{1983ApJ...272..317L}, to test the significance of the detected photon clusters. In the quadruplet case, ${\rm N}_{s}=$ 4 and ${\rm {N}_{B}}=$ 150638 (the photon number within the analyzed sky region centered at Position C). ${\rm t}_{s}$ is the time interval and ${\rm t}_{s}=$ 15.6 s. $\alpha$ is the ratio between the time interval and the total live time ${\rm t}_{b}$ (5.48 years). A significance of 6.1$\sigma$ is obtained for the photon quadruplets with ${\rm t}_{s}=$ 15.6 s. The photon quadruplet contains a 115-141 MeV photon triplet (${\rm N}_{s}=$ 3) with ${\rm t}_{s}=$ 5.0 s, which corresponds to a significance of 5.8$\sigma$ for Poisson fluctuation. In the general cases that photons were extracted within the 68\% PSF at the FRB position, the total photon number ${\rm {N}_{B}}=$ 105223. For a photon doublet with ${\rm t}_{s}<$ 2.0 s, a significance of $>$5.1$\sigma$ is obtained, while for photon triplets with ${\rm t}_{s}=$ 15.6 and 29.0 s, significance of 5.3$\sigma$ and 5.0$\sigma$ are obtained, respectively. 

The question we pursue is more than the significance of the flares as timing signals; our question is whether some of the photons are from the FRB source. Independent of the estimate of the significance of the short \gr\ flares as a timing signal from a source, there is additional spatial coincidence with \target\ in those \gr\ flares detected as short-duration photon clusters, which implies the associations are even more significant than those estimates based on photon clusters. The maximum likelihood analysis for the quadruplet has a TS value of $\sim$21. The best-fit position of the flare is only 0.3\arcdeg\ away from the precise position of \target, well within the 1$\sigma$ nominal uncertainty of 0.5\arcdeg. As the maximum likelihood analysis considers all the photons in the data no smaller than the 20$^{\circ}$ $\times$ 20$^{\circ}$ square region surrounding \target, the chance for a timing signal falls onto the position of \target\ by accident can be roughly estimated as less than ${0.5}\times{0.5} \pi$ out of ${20}\times{20}$ square degrees, which gives 0.002. The chance of spatial coincidence is more than enough to compensate considerations of the number of trials in our search for photon clusters. The total photon counts we extracted from the central PSF at the FRB position are ${\rm N}_{1}$=202 for E1, ${\rm N}_{2}$=288 for E2, and ${\rm N}_{3}$=151 for E3, respectively. At the Position C, ${\rm N}_{3c}$ = 217 photons for E3. A conservative estimate of the significance might take our searches for doublets, triplets, and quadruplets as having more number of trials of $\sim~{\rm N}_{i}$ where i=1,2,3. This will reduce the significance of the flare as a timing signal. 

Notice that any detected doublets, triplets and quadruplets are either due to an underlying source or due to background fluctuations, or both. First, if there is an underlying variable source having short-term variability,  we will detect more doublets, triplets and quadruplets, and so on, than pure Poison fluctuation. This is what we see in E1, E2, and E3 as inferred from Figure \ref{fig:li_ma}, suggesting a variable source is behind. Second, if the doublets, triplets, and quadruplets indeed originate from a variable source, they will reflect the source position. This is what we have seen: if, alternatively, those photon clusters were due to fluctuations, the average positions would simply not converge to the FRB position. In summary, we have found strong evidence for a temporal and spatial association of the \gr\ emission with \target.


\subsection{Theoretical models\\}

The key observations are a 15.6-s GeV \gr\ flare with luminosity $\sim 10^{48} \ {\rm erg \ s^{-1}}$ inferred from the quadruplet detected in Episode 3 and a few more possible \gr\ flares inferred from doublets or triplets detected in Episodes 1, 2 \& 3. The 5-day average luminosity from the source can reach as high as $\sim 10^{45} \ {\rm erg \ s^{-1}}$ as seen in E3. The total time span of these flares cover more than 3 months. These observations place constraints on the nature of the \gr\ flaring source and its connection with the FRB source. 

\begin{itemize}
\item The GeV luminosity of $\sim 10^{48} \ {\rm erg \ s^{-1}}$ rules out all known \gr\ sources other than GRBs and blazars, the only two types of \gr\ sources known to reach such luminosities. 
\item A GRB-FRB association is possible if a GRB is associated with the birth of the FRB engine. In this scenario, the GRB should lead all FRBs \cite{ZhangARNPS24}. However, the GeV flare was detected after the FRB source became active. So such a possibility is ruled out. Also, if one considers other potential \gr\ flares spanning a period of more than three months, the GRB association is ruled out because GRB emission typically decays within hours and should not reactivate.
\item A blazar origin of the \gr\ emission is possible. However, since no \gr\ emission was detected at the position in the past, this blazar should be dormant in the past and became active only recently. Also, the blazar should not be the source of FRB itself. This is because the minimum variability timescale of the supermassive black hole at the center of a blazar is defined by the light travel time of its horizon, which is $\delta t_{\rm min} \sim 2 G M / c^3 \sim 100 \ {\rm s} \ (M/10^7 M_\odot)$, which is much longer than the typical millisecond duration of FRBs. In order to make a millisecond-duration FRB, the black hole mass should be below $100 M_\odot$, i.e. it should belong to a stellar-mass black hole. On the other hand, such a blazar could serve as an external source to trigger FRBs from a FRB engine. 
\item 
Another possibility is a micro-blazar, a blazar-like system with the engine of a stellar-mass black hole or a neutron star. Some Galactic microquasars have been detected \cite{mirabel94}, but with much lower luminosities. The envisaged micro-blazars, if exist, must be much more luminous and probably with a much narrower beaming angle than known systems \cite{2023ApJ...959...89Z}.
\end{itemize}

We consider two possibilities to interpret the connection between the GeV \gr\ flares and \target:

\begin{enumerate}
\item {\bf The \gr\ flaring source is the source of the FRB. } \\
Within this scenario, the \gr\ source needs to be a stellar scale compact object because of the duration constraint of FRBs. We consider two possibilities:

 \begin{itemize}
 
  \item The source is a {\it newborn millisecond pulsar}. The total spin energy of a newborn millisecond pulsar is $E_{\rm spin} \simeq (1/2) I \Omega_0^2 \simeq (2\times 10^{52}) \ {\rm erg} \ (M/1.4 M_\odot) R_6^2 P_{0,-3}^{-2}$, where $I$, $M$, $R$, $\Omega_0$, and $P_0$ are the moment of inertia, mass, radius, initial angular velocity and period of the newborn magnetar, respectively, and the convention $Q_n=Q/10^n$ has been adopted in cgs units. In order to be active for $\tau > 3$ months, the spindown luminosity should be $L_{\rm sd} < E_{\rm spin} / \tau \simeq (2.3 \times 10^{45}) \ {\rm erg \ s^{-1}} R_6^2 P_{0,-3}^{-2}$. Because the pulsar spindown luminosity can be written as $L_{\rm sd} = B_p^2 R^6 \Omega_0^4 / 6c^3 \simeq 10^{47} \ {\rm erg \ s^{-1}} B_{p,14}^2 P_{0,-3}^{-4} R_6^6$, the upper limit of the spindown time sets an upper limit of the polar magnetic field strength: $B_p < (1.5 \times 10^{13} \ {\rm G}) P_0 R_6^{-2}$. Within such a scenario, this central engine is not a traditionally defined magnetar with $B_p > 10^{14} \ {\rm G}$, but is required to occasionally emit \gr\ flares with peak luminosities $>10^{48}~{\rm erg \ s^{-1}}$. This is not impossible, because a millisecond rotator may carry a large toroidal magnetic field due to differential rotation \cite{duncan92}, and when these strong magnetic field bubbles escape the neutron star surface, some \gr\ flares may be made \cite{kluzniak98,dai06}. FRBs may be made via magnetic flares \cite{lyubarsky14,metzger19,beloborodov20,lu20} or giant pulses \cite{cordes16}. One caveat of such a scenario is that the environment of a newborn pulsar from a supernova may be too dirty to allow free propagation of FRBs \cite{luan14}, but a GRB-like weak jet and a strong pulsar wind may blow a cavity to allow FRBs to escape \cite{zhang14,wang24}. 
  
  \item The source is a {\it newborn accreting stellar-mass black hole or a neutron star with a relativistic jet that produces \gr\ flares}. This scenario invokes a mechanism similar to microquasars in the galaxy \cite{mirabel94}, with the stellar-mass black hole or neutron star accreting from a binary companion and launching a relativistic jet. These objects can produce super-Eddington emission in the form of \gr emission, but the luminosities of the known Galactic microquasar sources are about 9 orders of magnitude smaller than our \gr\ flaring source. Thus, such a ``micro-blazar'' system, if exists, must involve a hitherto undetected extremely narrow relativistic jet pointing squarely toward Earth. 
  
\end{itemize}
 
\item {\bf The \gr\ flaring source is an external source that trigger FRBs.} \\
An alternative possibility is that the \gr\ flaring source itself is not the FRB engine, but it is just by chance in the vicinity of a regular FRB engine (for example, a magnetar). The intense \gr\ flares, as well as their associated strong relativistic outflows, would significantly impact the magnetosphere  of the FRB engine, causing FRB triggers \cite{zhang17}. Within this scenario, the \gr\ flaring source itself could be a reactivated blazar, because the FRB timescale is not defined by the blazar variability, but the variability of the magnetar engine \cite{zhang18}. One concern is how much role the \gr\ source is playing in triggering FRBs, as other active FRBs do not have GeV counterparts \cite{zhangzhang17,yang19} and seem not to require an external trigger.  There are two possibilities. First, \target\ is indeed more active than most other known repeating FRBs. The extra activity might benefit from the enhanced triggers induced by the \gr\ flaring. More likely, other active repeaters may indeed have enhanced triggers from a nearby object, which is not energetic enough or beamed favourably to be detected at cosmological distances. Such a possibility is supported by growing evidence of the possible existence of a companion in the vicinity of several active repeating FRBs \cite{ioka20,anna-thomas23}. 
\end{enumerate}

\newpage
\bigskip

\clearpage
\begin{table}
\centering
\setlength{\tabcolsep}{2pt}
\caption{List of the Photon Doublets Identified in Episode 1 and 2}
\label{tab:2p}
\begin{tabular}{cccccccc}
\hline
Time & $\mathrm{\Delta}t$ & RA & DEC & Energy \\
(MJD) & (s) & (Degree) & (Degree) & (MeV) \\
\hline
60334.209556 & 0.0 & 323.948 & 3.19857 & 262.369 \\
-- & 0.7 & 319.192 & 8.60994 & 132.747 \\
\hline
60369.142422 & 0.0 & 324.599 & 4.77732 & 266.840 \\
-- & 1.6 & 321.131 & 1.13108 & 277.715 \\
\hline
60373.092853 & 0.0 & 325.630 & 2.07277 & 167.152 \\
-- & 1.8 & 323.953 & 2.98876 & 120.644 \\
\hline
\end{tabular}
\vskip 1mm
\footnotesize{Note: These photon doublets correspond to photons collected surrounding the FRB.}
\end{table}

\clearpage
\begin{table}
\centering
\setlength{\tabcolsep}{2pt}
\caption{List of the Photon Triplets and Quadruplets Identified in Episode 3}
\label{tab:3p}
\begin{tabular}{cccccccc}
\hline
Time & $\mathrm{\Delta}t$ & RA & DEC & Energy \\
(MJD) & (s) & (Degree) & (Degree) & (MeV) \\
\hline
60428.845185 & 0.0 & 321.944 & 4.58652 & 192.174 \\
-- & 22.9 & 319.010 & 0.815437 & 107.400 \\
-- & 29.0 & 325.052 & 3.68540 & 271.049 \\
\hline
60429.843423 & 0.0 & 322.200 & 4.21218 & 1014.16 \\
(C) & 10.6 & 320.154 & -1.43609 & 137.720 \\
-- & 11.6 & 324.435 & 1.61275 & 115.717 \\
-- & 15.6 & 321.483 & 5.48386 & 141.357 \\
\hline
\end{tabular}
\vskip 1mm
\footnotesize{Note: The photon quadruplet corresponds to photons collected surrounding Point C -- the EAD position of the FRB, S1, and S2.}
\end{table}

\clearpage
\begin{figure}
   \includegraphics[height=0.45\textwidth]{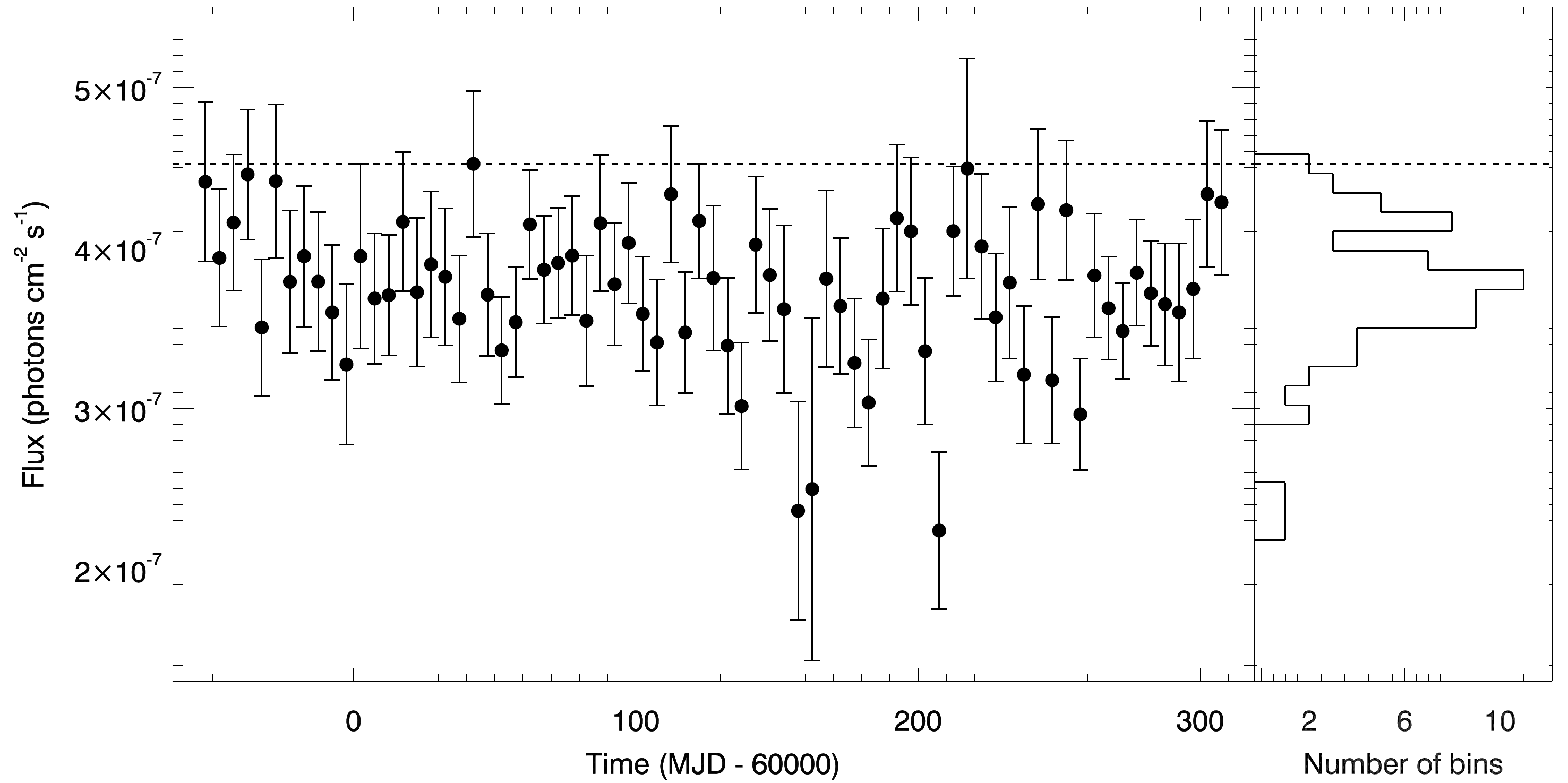}
   \caption{\textbf{The 0.1--500 GeV 5-day averaged LAT aperture photometry light curve during 2023 (MJD 59945--60310) and the corresponding histogram of number of bins.} The horizontal dashed line marks the maximum flux level that was reached during 2023.
}
   \label{fig:aplc}
\end{figure}

\clearpage
\begin{figure*}
\centering
\includegraphics[width=0.9\textwidth,angle=0]{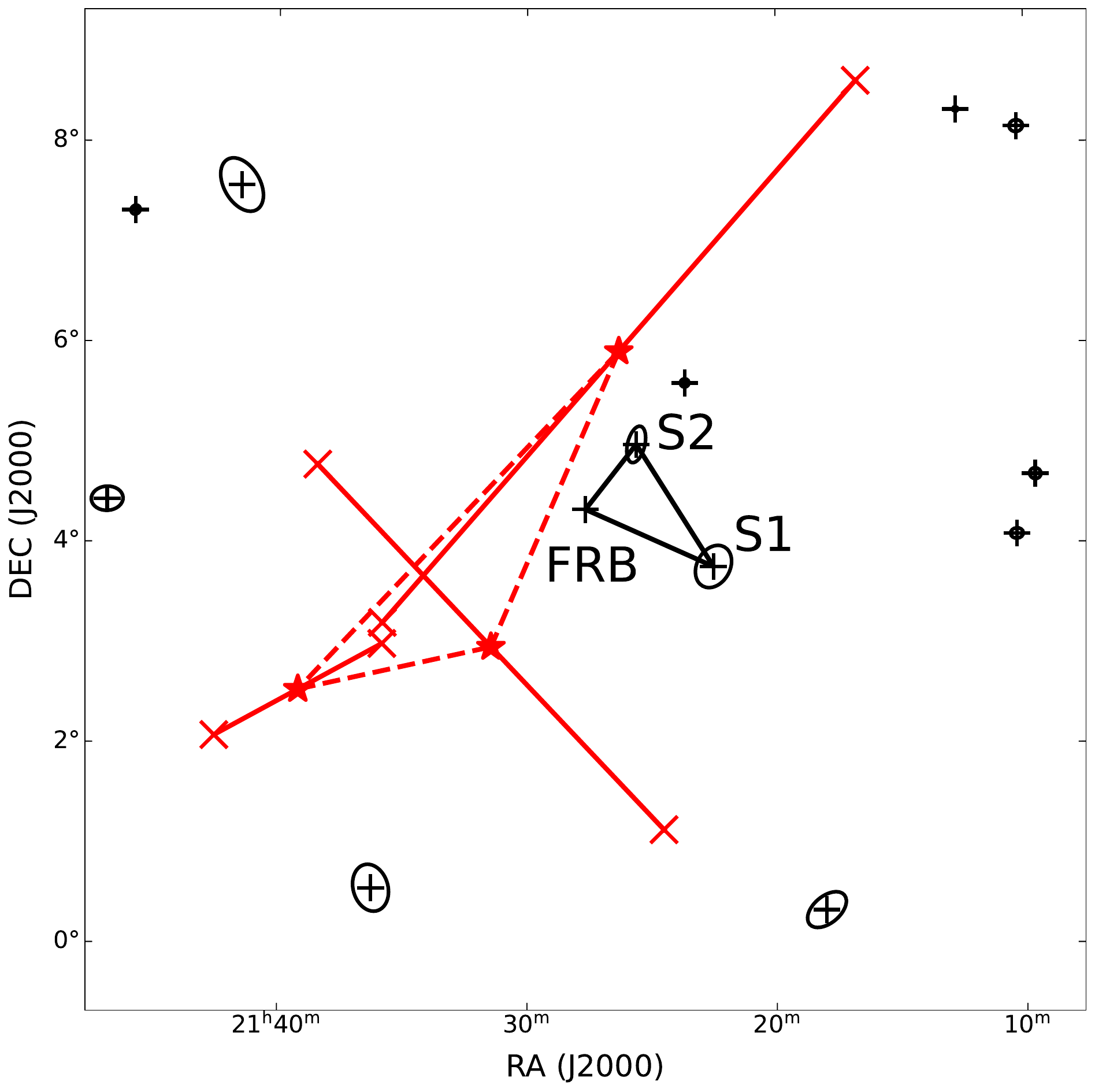}
   \caption{\textbf{The $10\arcdeg\times10\arcdeg$ FOV centerd at the FRB position (the 68\% PSF radius at 100 MeV is 5.3\arcdeg).} The black plus sign and ellipse are the positions and the 95\% error ellipses of sources given in the Fermi source catalogue. The black plus sign at the center marks the position of the FRB. The three pairs of photons (doublets) are marked by red crosses and are connected by a red solid line, with the midpoints of each doublet marked by a red star. The triangle formed by the three red stars (midpoints of doublets) implies the sky region favoured by the three doublets. The center of the red dashed triangle is on the lower-left side of the source positions of the FRB , S1 and S2, thus these doublets are in favour of the association with the FRB. 
}
   \label{fig:2p}
\end{figure*}

\clearpage
\begin{figure*}
\centering
\includegraphics[width=0.9\textwidth,angle=0]{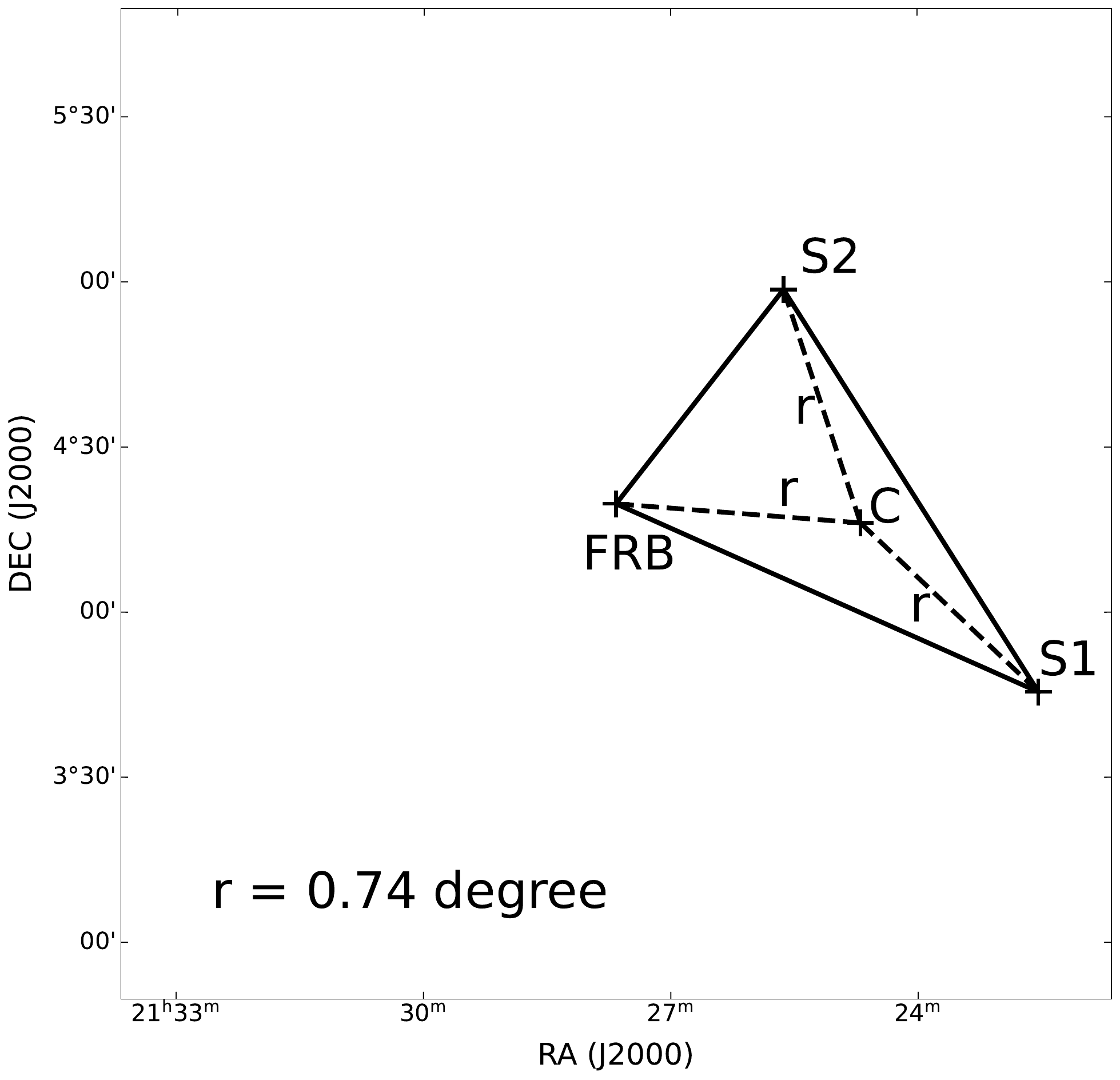}
   \caption{\textbf{The determination of the Equal Angular Distance (EAD) position C from the FRB, S1 and S2.} It's coordinates are R.A.=321.17\arcdeg, Decl.=4.27\arcdeg, with the equal angular separation r$=$0.74\arcdeg\ from S1, S2, and the FRB. Centered at position C, photon selection has no bias towards each of the three source positions. 
   }
   \label{fig:cpoint}
\end{figure*}

\begin{figure*}
\centering
\includegraphics[width=0.98\textwidth,angle=0]{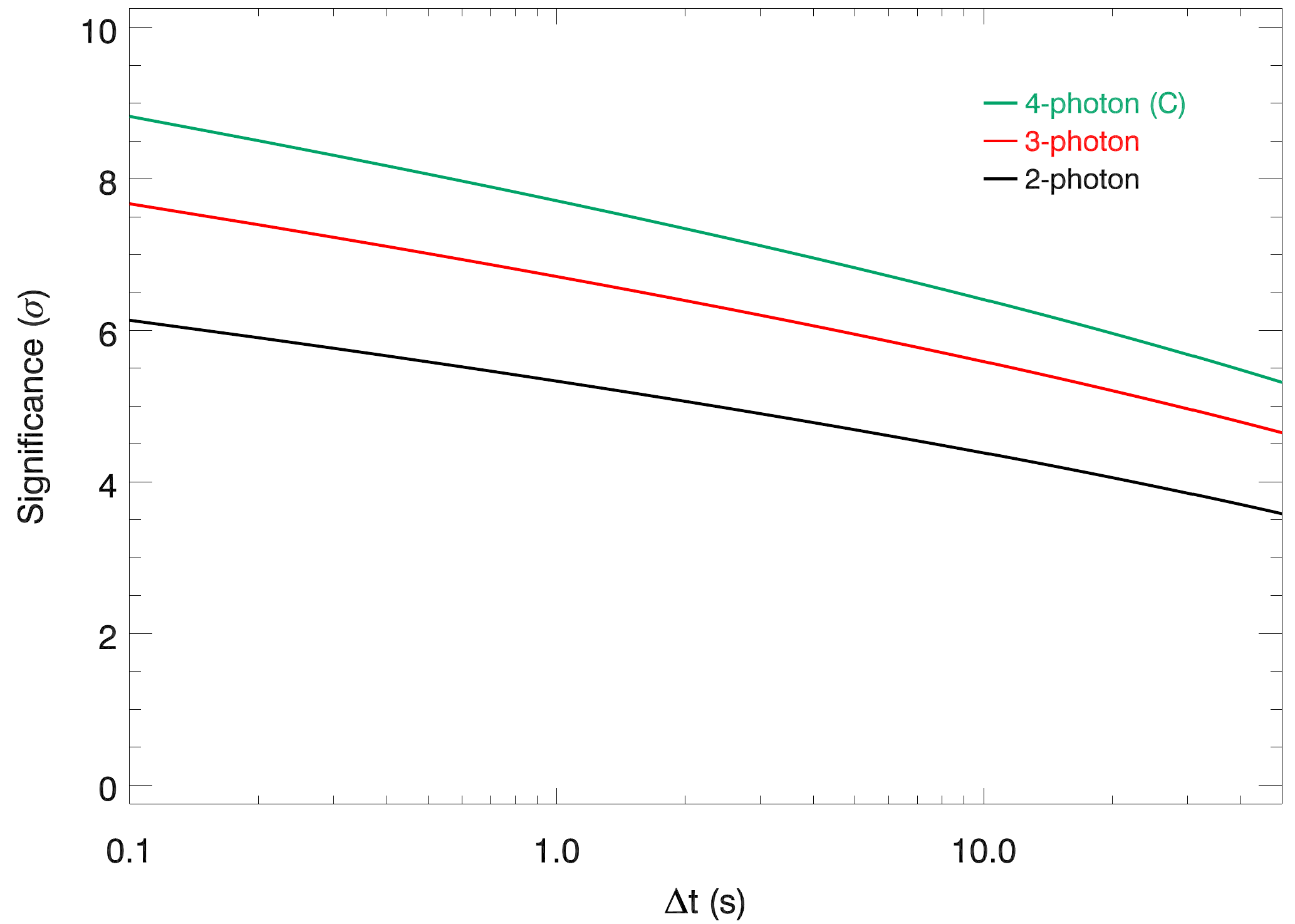}
\caption{
\textbf{ The theoretical relation between the significance of photon clusters and the time interval based on Poison fluctuation and the Fermi LAT observations (see \cite{1983ApJ...272..317L,2021NatAs...5..385F})}. The doublet and triplet results correspond to the FRB position, and the quadruple result corresponds to the EAD position C and a circular sky region with a radius ${\rm r}=0.74$ degree larger than the 68\% PSF. 
}
\label{fig:li_ma}
\end{figure*}

\end{methods}

\end{document}